\documentclass{epl} 
\newcommand{\be}{\begin{equation}}      
\newcommand{\ee}{\end{equation}}      
      
\newcommand{\bef}{\begin{figure}}      
\newcommand{\eef}{\end{figure}}      
\newcommand{\bea}{\begin{eqnarray}}    
\newcommand{\eea}{\end{eqnarray}}

\def\spose#1{\hbox to 0pt{#1\hss}}      
\def\ltapprox{\mathrel{\spose{\lower 3pt\hbox{$\mathchar"218$}}      
 \raise 2.0pt\hbox{$\mathchar"13C$}}}      
\def\gtapprox{\mathrel{\spose{\lower 3pt\hbox{$\mathchar"218$}}      
 \raise 2.0pt\hbox{$\mathchar"13E$}}}      
\def\inapprox{\mathrel{\spose{\lower 3pt\hbox{$\mathchar"218$}}      
 \raise 2.0pt\hbox{$\mathchar"232$}}}

\newcommand{\bean}{\begin{eqnarray*}}  
\newcommand{\eean}{\end{eqnarray*}}  
  
\def\lsim{\raise 0.4ex\hbox{$<$}\kern -0.8em\lower 0.62ex\hbox{$\sim$}}  
\def\gsim{\raise 0.4ex\hbox{$>$}\kern -0.7em\lower 0.62ex\hbox{$\sim$}}

\begin{document}  
\shorttitle{Causality constraints}

\title{Causality constraints on fluctuations in cosmology:
a study with exactly solvable one dimensional models}

\author{A.Gabrielli \inst{1}, M. Joyce\inst{2}, 
B. Marcos \inst{3} and P. Viot \inst{4}}
\institute{   
\inst{1}  ``E. Fermi'' Center, Via Panisperna 89 A, Compendio del 
Viminale, 00184 - Rome, Italy.\\
\inst{2} Laboratoire de Physique Nucl\'eaire et de Hautes Energies,  
 Universit\'e de Paris VI, 4, Place Jussieu, 
Tour 33 -RdC, 75252 Paris Cedex 05, France.\\
\inst{3} Laboratoire de Physique Th\'eorique,  
         Universit\'e de Paris XI, B\^atiment 211,  
        91405 Orsay, France.\\
\inst{4} Laboratoire de Physique Th\'eorique des Liquides,
4, Place Jussieu, 75252 Paris Cedex 05, France.\\  
}   
%
\maketitle

\begin{abstract}
A well known argument in cosmology
gives that the power spectrum (or structure function) 
$P(k)$ of mass density fluctuations 
produced from a uniform initial state by 
physics which is causal (i.e. moves matter and momentum only up 
to a finite scale) has the behaviour $P(k) \propto k^4$ at small $k$.
Noting the assumption of analyticity at $k=0$ of $P(k)$
in the standard derivation of this result, we introduce a class of 
solvable one dimensional models 
which allows us to study the relation between 
the behaviour of $P(k)$ at small $k$ 
and the properties of the probability  
distribution $f(l)$ for the spatial 
extent $l$ of mass and momentum conserving fluctuations. 
We find that the $k^4$ behaviour is obtained 
in the case that the first {\it six} moments of $f(l)$
are finite. Interestingly the condition that the 
fluctuations be localised - taken to correspond to
the convergence of the first two moments of $f(l)$ -
imposes only the weaker constraint $P(k) \propto k^n$  
with $n$ anywhere in the range $0< n \leq 4$. 
We interpret
this result to suggest that the causality bound will
be loosened in this way if quantum fluctuations 
are permitted.

\end{abstract}

In cosmology ``causality bounds'' are very important
in various contexts. They are limits which can be inferred 
simply from the fact, which is an intrinsic feature of Big Bang
models, that there is a finite horizon for causal processes i.e. 
light can travel only a finite distance in the time since
the Big Bang. One 
important example is an argument, due originally to Zeldovich
\cite{zeldovich1965}, which gives a strong constraint on the power
spectrum (i.e. what is usually called the structure function in
statistical physics) describing mass fluctuations. It states
that if fluctuations are built, starting from a uniform distribution
of matter, by causal physics (i.e. physical
processes moving matter and momentum coherently up to
a maximal scale), then the small $k$ form of the power
spectrum is $P(k) \propto k^4$.  The importance of this argument
is in its corollaries: it implies that the spectrum considered to 
correctly describe the perturbations at very large scales observed
\cite{cobe} in the microwave background, $P(k) \sim k$, cannot be 
produced by causal   physics acting prior  to the  time when radiation
decouples from matter.  And  it  is one of   the motivations  for  and
successes of the popular ``inflation''  model that it can produce such
fluctuations (by modifying the causal  structure of the Big Bang model
at  early times). In this letter  we describe a  study of this bound -
for which there    is  no rigourous   demonstration -   from  a purely
statistical  physics perspective: causality is  simply a  bound on the
distance  over which  matter  can  be moved coherently   in building a
fluctuating mass distribution from an initially uniform background. We
introduce a set of exactly solvable one dimensional models which allow
us to study  how  the $k^4$ result  depends on  the precise constraint
which is assumed to be imposed on real space fluctuations. Our central
result  is that the condition  of localisation of the fluctuations
(in the sense usual   in statistical physics  i.e.  finite   first and
second moment) leads only to the  weaker result $P(k) \propto k^n$ with $0<n
\leq 4$, with the $k^4$ result requiring additional constraints on higher
moments.  In   the context of cosmology  this  suggests that the $k^4$
bound should not be taken as valid when quantum fluctuations are taken
into account.

Let us consider first a simple derivation of this $k^4$
bound on fluctuations
\footnote{We note that condensed matter also provides examples
of systems where the structure factor shows this small $k$-behaviour.
In particular when a system  is quenched from  a
homogeneous (high   temperature)  phase into   a  broken-symmetry (low
temperature) phase, the  kinetics  of the system is   characterized by
order growth \cite{Bray02}, and the structure  factor associated to the
field  describing the order  has  at  late stages of this process  
a scaling   expression  $S(k,t)\sim k_m^{-3}F(k/k_m(t))$     
where  $k_m$ corresponds to the  maximum of $S$ at a given 
time  $t$. For systems whose phase   ordering  is  characterized by 
a  scalar field  which is conserved during the process, the  
scaling function $F(x)$  goes   as $x^4$\cite{Yeung,fratzl}.}.
The power spectrum $P(\vec{k})$ is
defined as $P(\vec{k})=\lim_{V\to\infty} (\left<|\delta_\rho
(\vec{k})|^2\right>/V)$ where 
\[\delta_\rho (\vec{k})=\int_V d^3x \, e^{-i\vec{k}\cdot \vec{x}}
\delta_\rho (\vec{x})\,.\]
and $\delta_\rho (\vec{x}) = (\rho(\vec{x})-\rho_0)/\rho_0$ is the
density fluctuation field around the mean density $\rho_0$.
Assuming statistical homogeneity $P(\vec{k})$
is also given by the Fourier transform
of the two point correlation function 
$\xi(\vec{x})=\left<\delta_\rho(\vec{x}_0)\delta_\rho(\vec{x}_0+
\vec{x})\right>$. Expanding $P(\vec{k})$ 
in powers of $\vec{k}$ one obtains,
assuming statistical isotropy,
\be
\begin{array}{l}
P(\vec{k})\equiv P(k)= \\
\int \xi(x) d^3x -\frac{k^2}{3!}\int x^2\xi(x) d^3 x +
\frac{k^4}{5!} \int x^4 \xi(x) d^3x
+ 0(k^6) 
\end{array}
\label{ps-expansion-in-k}
\ee
where the integrals are now over all space, 
$k=|\vec{k}|$ and $x=|\vec{x}|$. 
The $k^4$ result is obtained
if the zero and second moments of $\xi(x)$
vanish. Substituting a Taylor
expansion of $\delta_\rho (\vec{x})$ 
\footnote{In cosmology the standard
derivation of the bound (see e.g. 
\cite{pee80, padm, carr-silk, abbott-traschen})
uses directly such a Taylor expansion of $\delta_\rho (\vec{x})$, which is 
problematic as the coefficients in this expansion are formally divergent.}
to obtain $P(k)$, it is easy to show that
these two conditions are equivalent to the requirement
\bea
P(0)\!&=&\!\lim_{V\to\infty}
\frac{\langle (\Delta M)^2 (V) \rangle}{V} =0 
\label{condition-1}\\
P''(0)\!&\propto&\!\lim_{V\to\infty}
\frac{\langle \int_V 
(\vec{x}-\vec{y})^2 
\delta_\rho (\vec{x}) \delta_\rho (\vec{y})d^3 x d^3 y \rangle}{V}\!=\!0.
\label{condition-2}
\eea
That the first of these conditions Eq.(\ref{condition-1}) 
follows as a consequence of mass 
conservation and the causality constraint can be 
understood simply. For it is just equivalent to the 
condition that the fluctuations are sub-poissonian i.e. 
with a squared mass variance which grows less rapidly
than the volume. Poisson fluctuations are an
upper limit because in order to obtain them from 
an initially uniform distribution, one clearly
needs to move mass randomly over an average scale 
proportional to the size of the system. The causality 
constraint however places
an upper bound on this scale.

The  link  between  the second condition   Eq. (\ref{condition-2}) and
momentum conservation is less evident. Indeed the integral is not just
proportional  to  the  variance  in   the momentum  given  by   $\langle \int_V
\vec{x}.\vec{y}\delta_\rho(\vec{x})\delta_\rho(\vec{y})  d^3 x  d^3  y \rangle$, which,   in
fact, diverges  faster than the integration  volume $V$. The necessary
finiteness of the coefficient of $k^2$ (if it is finite) is ensured by
the cancelling pieces which come from the coefficients of the constant
and  $k^2$ term in the expansion  of $\delta_\rho(\vec{k})$ i.e. from the zero
and  {\it second} moments of the  fluctuation  field \cite{fratzl}.

Putting aside this remaining subtlety in the demonstration of the
$k^4$ result, we concentrate here on another aspect of it:
the implicit assumption which has been made that the Taylor 
expansion of $P(k)$ at $k=0$ is well defined (at least to order $k^4$).
What is the content of this assumption, and
is it justified here?  Fourier transform theory tells us 
that if $\xi(x)$ is rapidly decreasing at large $x$, then 
its Fourier transform $P(k)$ is analytic at $k=0$. 
In cosmology causality might be - and often is - taken
to require that physical quantities are strictly 
uncorrelated beyond the horizon scale i.e. that any 
correlation function 
is {\it identically} zero beyond this scale. 
In this case the assumption of analyticity is thus justified.
Such a constraint on the correlation function is, however, wholly 
classical. In quantum field theory causality is imposed as a 
constraint on the commutators of operators at 
space-like separations \cite{peskin}, and 
does not require that correlation functions vanish.
Our approach in this letter - through the study of
the models we will now describe - is to define the 
constraint from causality as one on the form 
of the permitted fluctuations in real space, and then   
to determine what this constraint implies about 
the behaviour of $P(k)$ as $k \to 0$.

\begin{figure}
\onefigure[width=12cm,angle=0]{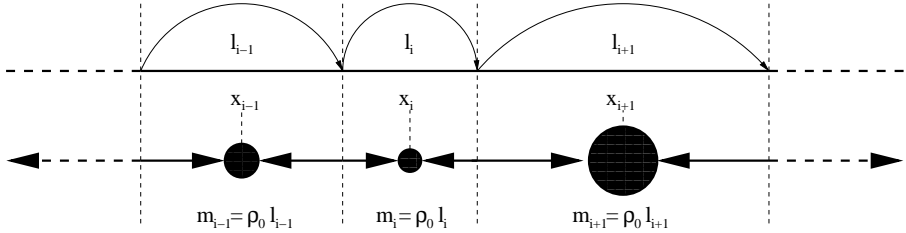} 
\caption{A schematic representation of our one dimensional models. Subsequent
uncorrelated steps of length $l$ with probability distribution $f(l)$ 
define a fragmentation of the line. The mass in each segment is then
aggregated to a point in its centre. Mass and centre of mass are thus 
explicitly conserved locally.} 
\label{fig1}  
\end{figure}

So let us consider now the following one dimensional
toy model for the generation of fluctuations
(see Fig.\ref{fig1}).
We start with a continuous exactly uniform mass
density $\rho_0$. Choosing randomly a first point, we 
take successive uncorrelated 
steps to the right of length $l$, 
where  $l$ is  a positive number chosen randomly 
at each step according to the probability density  $f(l)$. Doing
the same on the left, we define a division
of the real line into segments. 
We then gather
up the mass in each segment into a point, 
placing it in the {\it centre} of the segment. This defines
a mass and momentum conserving discretisation of
the initial mass density. The properties with respect 
to causality in this model depend wholly on those of 
the probability function $f(l)$,  as $l/2$ is the maximum 
distance through which mass and momentum must be moved 
to create the fluctuations. We will return after our
analysis of the model to discuss the subtleties of
this constraint. 
Note that the discreteness of the distribution
is not important here. By applying 
a smearing (e.g. gaussian) window function 
one could map the distribution onto a continuous
one. Such a smearing affects only the large
$k$ properties of $P(k)$, and thus
changes nothing for what concerns the results 
below.

We now derive exact expressions for the power spectrum 
in this model, and then study the small $k$ behaviour 
which interests us. The mass density in the segment 
$[0,L)$ of finite length $L$ is 
\begin{equation}
 \rho(x)=\rho_0 \sum_{m=1}^{N}l_m\delta(x-x_m)
\end{equation}
where 
$x_i=\sum_{j=1}^{i-1}l_j+l_i/2$
and the $l_i$ are the lengths of the $N$ consecutive segments
defining points in $[0,L)$. 
The finiteness of the first and second moments of $f(l)$
allows to us to apply  the central limit theorem to infer 
that the number of points $N$ 
in such a segment is gaussian distributed in the large $L$ limit about the
mean $\overline{N}=L/\langle l\rangle$ with fluctuations 
$\delta N\sim \sqrt{L}$.
The mean density is then defined i.e. we have
$\langle \rho(\vec{x}) \rangle= \rho_0$.    
From its definition it is easy to show that the 
power spectrum for $k\neq 0$ is given as  
$P(k)=\lim_{L\to \infty} P_L(k)$  
where
\begin{equation}
P_L(k)=\frac{1}{L} 
\left< \left|\sum_{m=1}^{N} l_me^{-ikx_m}\right|^2 \right>
\label{ps-sumform}
\end{equation}
which can be written as   
\begin{equation}
P_L (k)= \frac{1}{L} \left<  \sum_{i=1}^{N} l_i^2 \right>
+ \frac{2}{L} Re \left(\left<\sum_{i<j} l_i l_j e^{-ik(x_j-x_i)}\right>\right)
\,.
\end{equation}

Since we are interested in the limit
$L \to \infty$, we can evaluate the ensemble 
average with the number of points equal to the mean value
$\overline{N}$. The first ensemble average 
gives $\overline{N} \langle l^2 \rangle $, while the 
second, by using the definition of $x_i$, can be written as 
\be
\langle l e^{-ikl/2} \rangle^2 
\sum_{j=2}^{\overline{N}} \sum_{i=1}^{j-1} 
(\langle e^{-ikl}\rangle)^{j-i-1}  
\ee
where $\langle g(l) \rangle= \int_0^\infty dl g(l) f(l)$ 
for any function $g(l)$. Performing the sum we find the 
exact expression for the
power spectrum in the limit $L\to +\infty$
\begin{equation}
P(k)=
\frac{1}{\langle l \rangle} \left[\langle l^2 \rangle 
-2Re \frac{\left(\tilde{f}'(\frac{k}{2})\right)^2} {1-\tilde{f}(k)}
\right]
\label{resultPk}
\end{equation}
where 
$\tilde{f}(k)= \int_0^\infty e^{-ikl} f (l) dl$ 
is the characteristic function for the probability
distribution, and $\tilde{f}'({k})$ its first derivative.

We can now discriminate various different cases.
First let us suppose that $\tilde{f}(k)$ is an analytic
function at $k=0$. All the moments of the probability distribution
then exist and one can write the Taylor expansion
\be
 \tilde{f}(k)=\sum_{j=0}^{\infty}
\frac{(-i)^j}{j!}k^j \langle l^j \rangle\,.
\ee
Substituted in Eq. (\ref{resultPk}) this gives, to leading 
order at small $k$:
\begin{equation}
P(k)= \frac{k^4}{576}\left(\frac{\langle
l^2\rangle\langle l^3\rangle^2}{\langle l\rangle^3}+
\frac{\langle l^6\rangle}{\langle l \rangle }-
2\frac{\langle l^3\rangle \langle l^4\rangle}{ \langle l\rangle^2}\right) 
+ 0(k^6)
\label{result-analytic}
\end{equation}
reproducing the standard $k^4$ result, with a $P(k)$ which
is analytic at $k=0$. 

The derivation of Eq.~(\ref{resultPk}) required, however, 
only the convergence  of the first two moments.  
Let us suppose that $f(l)$
is such that only 
its first $n$ ($\geq 2$) moments
are finite i.e. that $f(l) \sim l_o^\alpha/l^{\alpha+1}$ for
large $l$, with $\alpha > 2$ (and therefore
$n$ the integer part of $\alpha$). In this case the small $k$ 
behaviour of $\tilde{f}(k)$ is given by 
\begin{equation}
 \tilde{f}(k)=\sum_{j=0}^{n}
\frac{(-i)^j}{j!}k^j \langle l^j \rangle
+A  k^\alpha l_o^\alpha +0(k^{n+1}) 
\label{smallk-nonanal}
\end{equation}
for non-integer $\alpha$, where 
\be
A=\int_0^{\infty} \frac{\mathrm{d} u}{u^{\alpha+1}}
\left( e^{-iu} - \sum_{j=0}^{n} \frac{(-i)^j}{j!}u^j \right)
\label{A-explicit}
\ee
For integer $\alpha$ there are logarithmic corrections
to these formulae. Substituting Eq. (\ref{smallk-nonanal}) 
in Eq. (\ref{resultPk}) we find the leading
small $k$ behaviour
\be  
P(k)= 
C_\alpha \frac{l_o^\alpha}{\langle l \rangle} k^{\alpha-2} + 0(k^{\alpha-1})
\label{smallk-nonanal-PS}
\ee
for $2< n \leq  6$, where $C_\alpha$ is a (positive) dimensionless 
constant proportional to the real part of $A$.
For $\infty> n >6$, when at least the first six moments
are defined, 
Eq. (\ref{result-analytic}) again is obtained, but with higher order
corrections which are non-analytic. The non-analyticity of $P(k)$ 
at $k=0$ implies that the
correlation function $\xi(x)$, which in the case that all moments of
$f(l)$ are finite is rapidly decaying at large separations, shows
instead an algebraic tail ($\xi(x) \propto 1/x^{\alpha-1}$).

\begin{figure}
\onefigure[scale=0.6,angle=0]{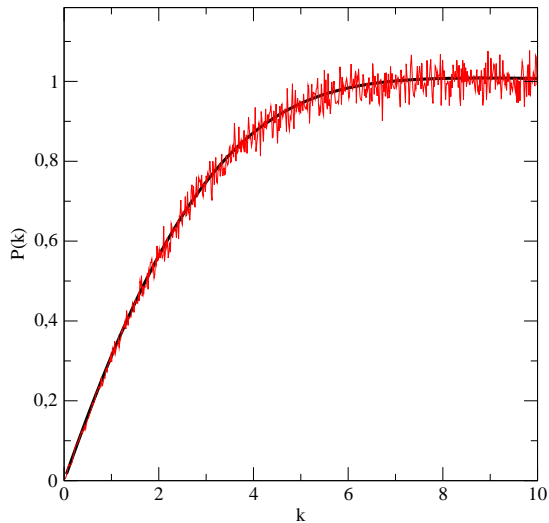} 
\caption{The power spectrum measured in 
a simulation of our one dimensional model with
$f(l)= (1+l^2)^{-2}$ (i.e. $\alpha=3$) for a box size $L=400$, and 
one thousand realisations. The black line is the exact
result obtained directly from Eq. (\ref{resultPk}). 
Both curves show the predicted
small $k$ behaviour $P(k) \propto k$.} 
\label{fig2}  
\end{figure}
While it is the asymptotic behaviour at small $k$ which we concentrate
on here, the expression Eq. (\ref{resultPk}) gives an exact expression
for the power spectrum for all $k$.
In Fig.~\ref{fig2} is shown both the result of such an exact
calculation for the case $f(l)= (1+l^2)^{-2}$ (i.e. $\alpha=3$), and
a numerical simulation of the same model. The expected small $k$
behaviour $P(k) \propto k$ can be clearly observed.
 
Let us consider the interpretation of these results.
We have been able to derive the small $k$ behaviour of
the power spectrum when the first two moments of $f(l)$
are taken to be finite. This requirement of the 
fluctuations corresponds simply to the standard
definition of localised fluctuations \cite{bouchaud}.
They are short-range correlations with a well
defined characteristic scale, in contrast to
long-range correlations (and so called
 ``broad'' distributions \cite{bouchaud}) which
lack such a scale. Thus we have imposed on the
fluctuations that they are effectively cut-off
at a finite length scale, which naturally then
explains the bound obtained: as noted earlier
the condition $P(0)=0$ corresponds exactly to
the requirement that fluctuations in
mass in real space should be sub-Poissonian.
When we move mass coherently only up to a 
finite scale, we expect Poissonian fluctuations
to form an upper bound on what one can obtain.
Note that this is also a result which we expect 
to be true independently of the spatial 
dimension, as the relation between the exponents of
the power spectrum and the classification (see \cite{gjsl}
for a discussion) of mass fluctuations in 
poissonian, sub-poissonian and super-poissonian 
is independent of dimension \footnote{A generalization
of this model to arbitary dimensions will be presented
elsewhere\cite{gjmv-prep}.}. Note also that the correlation
function, as we have seen, decays at least as fast as a
power-law with exponent greater than the dimension
($\alpha>2$), which indeed corresponds to an
effectively short-range
correlation with an integrable correlation function
(whose integral is zero). 

While we have this simple interpretation of our results,
what we have found is actually very non-trivial and 
surprising in another respect: the exponent at small $k$,
which characterises the large scale fluctuations {\it is not
universal} for this class of localised distributions
with finite mean and variance. Rather the exponent
depends in a subtle way on the convergence of higher
moments. We stress that this means the small-k behaviour
is thus determined by the large masses (i.e. the tails)
of the these distributions despite the fact that they
are not broad.

Now let us return to the interpretation of these results
in relation to the cosmological causality bound, which was
our original motivation. The problem of the causality
bound can be addressed with this simple model, because 
$l/2$ is the maximum distance through which mass and 
momentum must be moved to create the fluctuations.
If we impose that $f(l)$ is {\it strictly} zero above 
some finite scale $l_c$, no mass can be moved over
a scale larger than this ($l_c$ is then the ``size
of the causal horizon''). Such a constraint is what
one would take to be imposed classically by causality.
And our model gives in this case the $k^4$ result,
in accord with the standard cosmological bound.
However our model shows that a generalisation
of this result to the case where the fluctuations
are cut off at a characteristic scale $l_c$, but
smoothly, is non-trivial. And it is precisely
such a generalisation which is needed if
the $k^4$ bound is to extended to the case in which quantum
fluctuations are admitted (i.e. that the physics
generating the fluctuations is quantum in nature).
This is so because quantum correlation functions 
generically do not vanish at any separation,
as the required sharp localisation of states is impossible
(cf. Reeh-Schlieder theorem \cite{streater-wightman}).
This means that to extend the standard result
also to include quantum fluctuations, one must assume that 
any (causal) quantum mechanical evolution giving rise to
fluctuations in an expanding
FRW Universe (with sub-luminal, non-inflationary expansion)
gives rise to, for example, a strictly rapid decay of 
correlation functions at super-horizon scales, and excludes 
the rapid power-law decay admitted by a generic definition 
of localisation of fluctuations. While a rapid decay as an
 exponential 
is indeed typical of purely quantum fluctuations in the 
simplest cases one can envisage (e.g. the quantum correlations
at space-like separations of a free massive scalar field 
in its ground state) such an assumption is too restrictive
a form of the condition 
that quantum fluctuations be localised. An interesting
analogy is given by screening of electrostatic forces 
in a neutral plasma, which is 
analagous to the effect of the existence of 
a causal horizon relative to a theory in flat space,
as it leads to a physical  
cut-off scale beyond which the effect of a source is
no longer felt. While such screening gives rise to
exponential decay of the (charge-charge) correlation 
function in the case of a Coulomb potential, it can be 
shown easily \cite{gjjlpsl03} that,  
for a $1/r^2$ potential, the same physical mechanism 
leads to an effective short-range potential which 
decays algebraically (as $-1/r^4$ in three dimensions). 
Indeed the spectrum of fluctuations (at thermal equilibrium)
at small $k$ in this case behaves as $P(k) \propto k$, instead of 
$P(k) \propto k^2$ in the case of electrostatics \cite{gjjlpsl03}.
This suggests that a power spectrum of fluctuations 
$P(k) \propto k$ at small $k$ might, for example, be obtained 
from an appropriate interacting field quantized in 
a (sub-luminally expanding) FRW background. Such a model
would provide an interesting alternative to the standard
(non-causal) inflationary mechanism of producing such
fluctuations at superhorizon scales. 

{\bf Acknowledgements}
We thank B. Jancovici, J.L.Lebowitz, J. Mourad, F. Sylos Labini and 
F. van Wijland for useful discussions and comments. A. G. thanks 
the Physics Department of the
University ``La Sapienza'' of Rome (Italy) for supporting this
research.

\end{document}